\documentclass[letter]{aa}
\usepackage{txfonts}
\usepackage{graphicx}

\begin{document}

\title{The faint end of the 250 $\mu$m luminosity function at $z<0.5$}

\author{L.~Wang\inst{1,2,3}, P.~Norberg\inst{3}, M.~Bethermin\inst{4}, N.~Bourne\inst{5}, A. ~Cooray\inst{6}, W.~Cowley\inst{3}, L.~Dunne\inst{5,7},  S.~Dye\inst{8}, S.~Eales\inst{7}, D.~Farrah\inst{9}, C.~Lacey\inst{3}, J.~Loveday\inst{10}, S.~Maddox\inst{5,7}, S.~Oliver\inst{10}, M.~Viero\inst{11}}

\institute{SRON Netherlands Institute for Space Research, Landleven 12, 9747 AD, Groningen, The Netherlands \email{l.wang@sron.nl} 
\and Kapteyn Astronomical Institute, University of Groningen, Postbus 800, 9700 AV Groningen, the Netherlands
\and ICC \& CEA, Department of Physics, Durham University, Durham, DH1 3LE, UK 
\and European Southern Observatory, Karl Schwarzschild Stra\ss e 2, 85748 Garching, Germany
\and Institute for Astronomy, University of Edinburgh, Royal Observatory, Edinburgh EH9 3HJ, UK 
\and Center for Cosmology, Department of Physics and Astronomy, University of California, Irvine, CA 92697, USA
\and School of Physics and Astronomy, Cardiff University, The Parade, Cardiff CF24 3AA, UK
\and School of Physics and Astronomy, University of Nottingham, University Park, Nottingham, NG7 2RD, UK
\and Department of Physics, Virginia Tech, Blacksburg, VA 24061, USA
\and Astronomy Centre, University of Sussex, Falmer, Brighton BN1 9QH, UK
\and Kavli Institute for Particle Astrophysics and Cosmology, Stanford University, 382 Via Pueblo Mall, Stanford, CA 94305, USA}

\date{Received / Accepted}

\abstract
   {}
   {We aim to study the 250 $\mu$m luminosity function (LF) down to much fainter luminosities than achieved by previous efforts.}
   {We developed a modified stacking method to reconstruct the 250 $\mu$m LF using optically selected galaxies from the SDSS survey and {\it Herschel} maps of the GAMA equatorial fields and Stripe 82. Our stacking method not only recovers the mean 250 $\mu$m luminosities of galaxies that are too faint to be individually detected, but also their underlying distribution functions. }
   {We find very good agreement with previous measurements in the overlapping luminosity range. More importantly, we are able to derive the LF down to much fainter luminosities ($\sim25$ times fainter) than achieved by previous studies. We find strong positive luminosity evolution $L^*_{250}(z)\propto(1+z)^{4.89\pm1.07}$ and moderate negative density evolution $\Phi^*_{250}(z)\propto(1+z)^{-1.02\pm0.54}$ over the redshift range $0.02 < z< 0.5$.}
   {}


\titlerunning{The faint end of the 250 $\mu$m luminosity function}

\authorrunning{Wang et al.}

\maketitle

\section{Introduction}

Luminosity functions (LF) are fundamental properties of the observed galaxy populations that provide important constraints on models of galaxy formation and evolution (e.g. Lacey et al. 2015; Schaye et al. 2015). Studying the LF  at far-infrared (FIR) and sub-millimetre (sub-mm) wavelengths is critical. Half of the energy ever emitted by galaxies has been absorbed by dust and re-radiated in the FIR and sub-mm (Hauser \& Dwek 2001; Dole et al. 2006). The spectra of most IR luminous galaxies peak in the FIR and sub-mm (Symeonidis et al. 2013; Casey et al. 2014).  Finally, our knowledge of the FIR and sub-mm LF  is relatively poor.

The first 250 $\mu$m LF measurement was made by Eales et al. (2009) with observations conducted using the Balloon-borne Large Aperture Submm Telescope (BLAST; Devlin et al. 2009). {\it Herschel} (Pilbratt et al. 2010) significantly improved over BLAST with increased sensitivity, higher resolution, and larger areal coverage.  Dye et al. (2010) detected strong evolution in the 250 $\mu$m LF out to $z\sim0.5$, using the {\it Herschel}-Astrophysical Terahertz Large Area Survey (H-ATLAS; Eales et al. 2010). Using the {\it Herschel} Multi-tiered Extragalactic Survey (HerMES; Oliver et al. 2012), Vaccari et al. (2010) presented the first constraints on the 250, 350, and 500 $\mu$m as well as the infrared bolometric (8-1000 $\mu$m) LF at $z<0.2$.  More recently, combining {\it Herschel} data with multi-wavelength datasets, Marchetti et al. (2016) derived the LF at 250, 350, and 500 $\mu$m as well as the bolometric LF over $0.02<z<0.5$. Evolution in luminosity ($L^*_{250}\propto(1+z)^{5.3\pm0.2}$) and density ($\Phi^*_{250}\propto(1+z)^{-0.6\pm0.4}$) are found at $z<0.2$. Marchetti et al. (2016), however, were  unable to constrain evolution beyond $z\sim0.2,$ as only the brightest galaxies can be individually detected at higher redshifts. Despite the significant progress  made, the determination of the LF is still hampered by many difficulties. Large samples over large areas are required for accuracy. We need to focus on smaller areas with increased sensitivity, however, to probe the faint end. At the {\it Herschel}-SPIRE (Griffin et al. 2010) wavelengths, confusion (related to the relatively poor angular resolution) is a serious challenge for source extraction, flux estimation, and cross-identification with sources detected at other wavelengths. In addition, issues such as completeness and selection effects due to the combination of several surveys are extremely difficult to quantify (e.g. Casey et al. 2012). 

In this paper, we present a new analysis of the 250 $\mu$m LF by stacking deep optically selected galaxy catalogues from the Sloan Digital Sky Survey (SDSS) on the SPIRE 250 $\mu$m images. We bypass some major difficulties in previous measurements (e.g. complicated selection effects, reliability of the cross-identification).  The paper is organised as follows. In Section 2, we describe the relevant data products from the SDSS and {\it Herschel} surveys.  In Section 3, we explain our stacking method, which recovers the mean properties and underlying distribution functions. In Section 4, we present our results and compare with previous measurements. Finally, we give conclusions in Section 5. We assume $\Omega_m=0.25$,  $\Omega_{\Lambda}=0.75$,  and $H_0=73$ km s$^{-1}$ Mpc$^{-1}$. Flux densities are corrected for Galactic extinction (Schlegel, Finkbeiner \& Davis 1998).

\section{Data}

\subsection{Optical galaxy samples from SDSS}

The SDSS Data Release 12 (DR 12) contains observations from 1998 to 2014 over a third of the sky (Alam et al. 2015) in $ugriz$. The DR 12 includes photometric redshift ($z_{\rm phot}$) using an empirical method known as a kd-tree nearest neighbour fit (KF) (Csabai et al. 2007), which is extended with a template-fitting method to derive parameters, such as $k$ corrections and absolute magnitudes, using spectral templates from Dobos et al. (2012). The DR 12 features an expanded training set (extending to $z=0.8$), an updated method of template-fitting, and a more detailed approach to errors (Beck et al. 2016). Following recommendations on the SDSS website, we selected galaxies (located in the three Galaxy And Mass Assembly (GAMA) equatorial fields with {\it Herschel} coverage) with photoErrorClass equal to 1, -1, 2, and 3, which have an average RMS error in ($1+z$) of 0.02, 0.03, 0.03 and 0.03, respectively. We constructed volume-limited samples in five redshift bins, $z1=[0.02, 0.1]$, $z2=[0.1, 0.2]$, $z3=[0.2, 0.3]$, $z4=[0.3, 0.4]$ and $z5=[0.4, 0.5]$. In each bin, we only selected galaxies that were bright enough to be seen throughout the corresponding volume, given the apparent magnitude limit is $r=20.4$ which corresponds to the 90\% completeness limit for single pass images (Annis et al. 2014). We also take the most adverse $k$ correction in a given redshift bin into account in deriving the luminosity limit owing to the nature of flux-limited surveys.

The SDSS stripe along the celestial equator in the south Galactic cap, known as "Stripe 82", was the subject of repeated imaging. The resulting depths are roughly 2 magnitudes deeper than the single-epoch imaging. We used the Stripe 82 Coadd photometric redshift catalogue constructed using artificial neural network (Reis et al. 2012). The median photo-$z$ error is $\sigma_z=0.031$ and the photo-$z$ is well measured up to $z\sim0.8$. Following the procedure applied to the DR 12, we also constructed volume-limited samples in five redshift bins.We performed $k$ corrections in the optical bands to $z=0.1$  using KCORRECT v4\_2 (Blanton et al. 2002; Blanton \& Roweis 2007). The luminosity limit as a function of redshift is calculated using an apparent magnitude limit of $r=22.4,$ which corresponds to the 90\% completeness limit for the Coadd data (Annis et al. 2014) . This deeper catalogue allows us to probe 250 $\mu$m LF down to even fainter luminosities than the DR 12 catalogue.

\begin{figure}
\centering
\includegraphics[height=2.5in,width=3.3in]{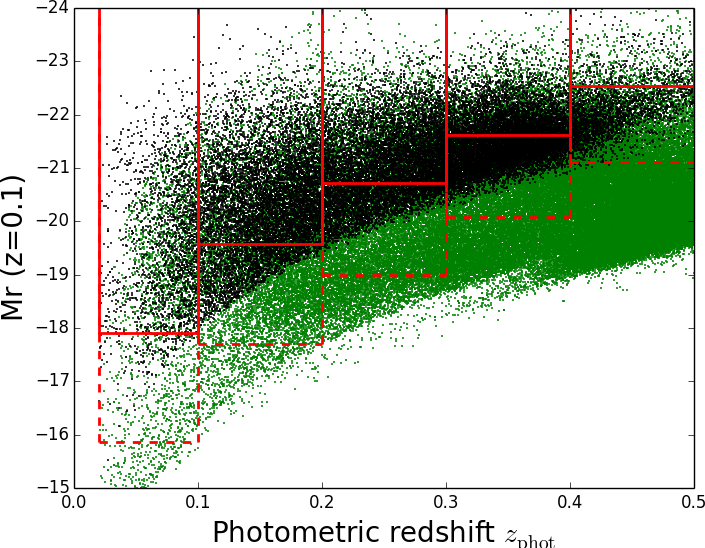}
\caption{Rest-frame r-band absolute magnitude $M_r$ vs. photometric redshift $z_{\rm phot}$ for DR12 galaxies with $r<20.4$ (black dots) and Stripe 82 galaxies with $r<22.4$ (green  dots), in areas with {\it Herschel}-SPIRE coverage. For clarity, only 20\% of the DR12 sample and $10\%$ of the Stripe 82 sample are plotted. The red boxes indicate the volume-limited subsamples in five redshift slices (solid: DR12; dashed: Stripe 82).}
\label{samples}
\end{figure}

Fig.~\ref{samples} shows the rest-frame $r$-band absolute magnitude $M_r$ ($k-$corrected to $z=0.1$) as a function of $z_{\rm phot}$ for galaxies with $r<20.4$ in the GAMA fields and for galaxies with $r<22.4$ in the Stripe 82 area with {\it Herschel} coverage. The red boxes indicate the redshift boundaries and $M_r$ limits used to define volume-limited samples. When carrying out the stacking procedure, we further bin galaxies in each redshift slice along the $M_r$ axis. The minimum bin width along $M_r$ is 0.15 mag but can be increased to ensure that the minimum number of galaxies in a given redshift and $M_r$ bin is 1000. 

\subsection{ {\it Herschel} survey 250 $\mu$m maps }

The H-ATLAS survey conducted observations at 100, 160, 250, 350, and 500 $\mu$m of the three equatorial fields also observed in the GAMA spectroscopic survey (Driver et al. 2011); these equatorial fields are G09, G12, and G15 centred at a right ascension of $\sim$9, 12, and 15 hours, respectively. For this study, we cut out a rectangle inside each of the GAMA fields with a total area of 95.6 deg$^2$. The version of the data used in this paper is the Phase 1 version 3 internal data release. The SPIRE maps, which have unit of Jy/beam, were made using the methods described by Valiante et al. (2016, in prep). Large-scale structures and artefacts are removed by running the NEBULISER routine developed by Irwin (2010). We estimated the local background by fitting a Gaussian to the peak of the histogram of pixel values in $30\times30$ pixel boxes and subtracted this background from the raw map.

As the deeper SDSS Coadd catalogue is located in Stripe 82,  we also used maps from the two {\it Herschel} surveys in the Stripe 82 region, i.e. the {\it Herschel} Stripe 82 Survey (HerS; Viero et al. 2014) and the HerMES Large-Mode Survey (HeLMS; Oliver et al. 2012). The joint HeRS and HeLMS areal coverage between -10$^{\circ}$ and 37$^{\circ}$ (RA) covers the subset of Stripe 82 that has the lowest level of Galactic dust emission (or cirrus) foregrounds. For this study, we combined 39.1 deg$^2$ in HeRS and 47.6 deg$^2$ in HeLMS, which are covered by the SDSS Coadd data. The SPIRE data, obtained from the {\it Herschel} Science Archive, were reduced using the standard ESA software and the custom-made software package, SMAP (Levenson et al. 2010; Viero et al. 2014). Maps were made using an updated version of SMAP/SHIM, which is an iterative map-maker designed to optimally separate large-scale noise from signal. Viero et al. (2013) provide greater detail on these maps.

\section{Method}

\begin{figure}
\includegraphics[height=2.in,width=3.4in]{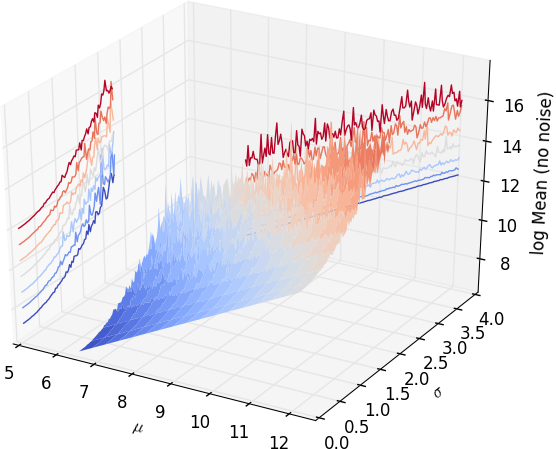}
\includegraphics[height=2.in,width=3.4in]{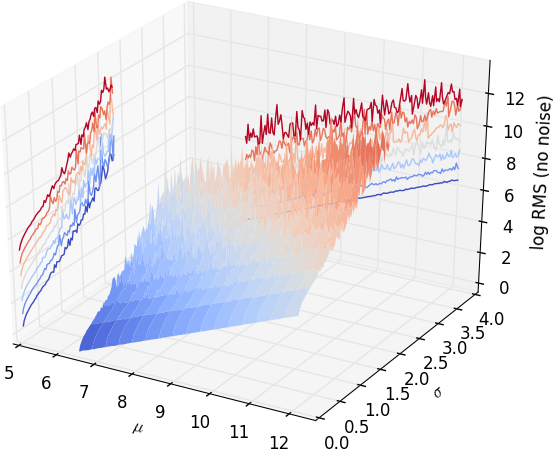}
\caption{Top: estimated mean $L250$ as a function of the intrinsic population mean ($\mu$) and standard deviation ($\sigma$) of $\log{L250}$. For each set of ($\mu$, $\sigma$), we generate $\sim$2000 random numbers representing the 250 $\mu$m luminosities drawn from the log-normal distribution specified by ($\mu$, $\sigma$). The estimates of the mean 250 $\mu$m luminosity $\bar{m}$ are derived from these specific realisations of log-normal distributions.  Bottom: the estimated standard deviation of $L250$ as a function of $\mu$ and $\sigma$.}
\label{method}
\end{figure}

Stacking was used for determining the mean properties  of sources detected at another wavelength that are individually too dim to be detected at the working wavelength.  For a given galaxy sample, we can stack\footnote{We use the IAS library (\url{http://www.ias.u-psud.fr/irgalaxies/files/ias_stacking_lib.tgz}) (Bavouzet 2008; B{\'e}thermin et al. 2010) to perform stacking. To avoid introducing bias, we did not clean the image of any detected sources. } the 250 $\mu$m images centred at the positions of the galaxies weighted by luminosity distance squared ($D_L^2$) and $k$ correction to derive the mean rest-frame 250 $\mu$m luminosity. To apply the $k$ correction at rest-frame 250 $\mu$m, we used
\begin{equation}
K(z) = \left(\frac{\nu_o}{\nu_e}\right)^{3+\beta} \frac{e^{h\nu_e/kT_{\rm dust}} - 1}{e^{h\nu_o/kT_{\rm dust} } - 1},
\end{equation}
where $\nu_o$ is the observed frequency and $\nu_e=(1+z)\nu_o$ is the emitted frequency in the rest frame. We assumed a mean dust temperature of $T_{\rm dust} = 18.5K$ and emissivity index $\beta=2$, following Bourne et al. (2012).

In this paper, we extend the traditional stacking method to reconstruct the LF. The key assumption is that the rest-frame 250 $\mu$m luminosities $L_{250}$ of galaxies in a narrow bin of $z$ and $M_r$ follow a log-normal distribution, i.e. the logarithm of the luminosities, $\log L_{250}$, follow a normal distribution with mean $\mu$ and standard deviation $\sigma$. In contrast, we used to $m$ denote the mean of $L_{250}$ and $s$ to denote the standard deviation of $L_{250}$. The two sets of parameters can be related to each other as,
\begin{equation}
\mu = \ln{ (m/\sqrt{1+s^2/m^2} )},  \sigma=\sqrt{\ln{(1 + s^2/m^2)}}.
\end{equation}
With stacking, we can estimate the mean  of $L_{250}$ ($\bar{m}$) and the standard deviation of $L_{250}$ ($\bar{s}$).  We use $m$ and $s$ to denote the intrinsic population mean and standard deviation parameters, and $\bar{m}$ and $\bar{s}$ to denote estimates\footnote{An estimator is a statistic, which is a function of the values in a given sample, used to estimate a population parameter. An estimate is a specific value of the estimator calculated from a particular sample. } of the intrinsic parameters.

To recover the LF in a given redshift bin, we need to infer $\mu$ and $\sigma$ as a function of $M_r$, using combinations of $\bar{m}$ and $\bar{s}$. In Fig. ~\ref{method}, we plot the estimated mean and standard deviation of $L_{250}$, i.e. $\bar{m}$ and $\bar{s}$ as a function of the intrinsic population mean and standard deviation of  $\log L_{250}$, i.e. $\mu$ and $\sigma$. To make this plot, we generated $\sim$2000 random numbers (representing the 250 $\mu$m luminosities) drawn from a log-normal distribution for each set of ($\mu$, $\sigma$) values. The estimates $\bar{m}$ and $\bar{s}$ were derived from these specific samples (i.e. realisations) of log-normal distributions. The estimates $\bar{m}$ and $\bar{s}$ become noisy when $\sigma$ is large (even in the absence of noise), even though $m$ and $s$ can be related to $\mu$ and $\sigma$ analytically (Eq. 2). This is because $\bar{m}$ and $\bar{s}$ are sensitive to the large values in the tail of the distribution. To take the effect of realistic noise into account, we injected synthetic galaxies with log-normally distributed $L_{250}$ (drawn from distributions of  known $\mu$ and $\sigma$) at random locations in the map. We can then  measure the mean and standard deviation of $L_{250}$  from the stacks of synthetic galaxies in the presence of realistic noise and compare with the estimated mean and standard deviation of $L_{250}$ from the stacks of real galaxies. We summarise the main steps of recovering the 250 $\mu$m LF using our modified stacking method in Appendix A.

\begin{figure}
\includegraphics[height=2.6in,width=3.4in]{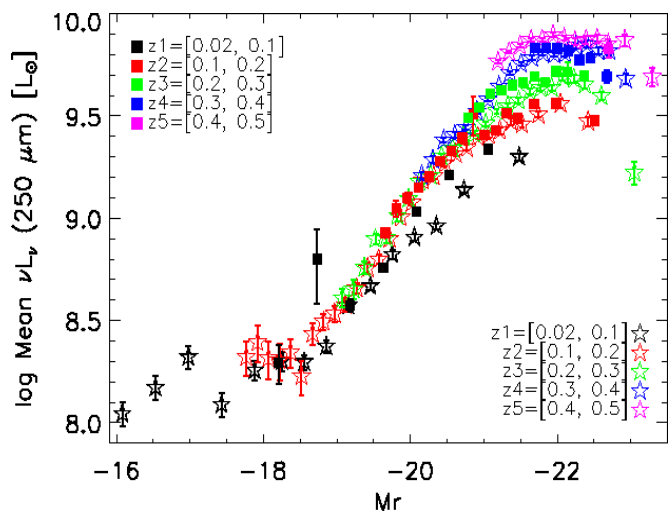}
\caption{Mean rest-frame 250 $\mu$m luminosity $L250$ vs. $Mr$ for the DR12 (solid squares) and Stripe 82 galaxies (open stars) in five redshift bins. Error bars correspond to the error on the mean.}
\label{mean_property}
\end{figure}

\section{Results}

\begin{figure}{\centering}
\includegraphics[height=3.9in,width=3.4in]{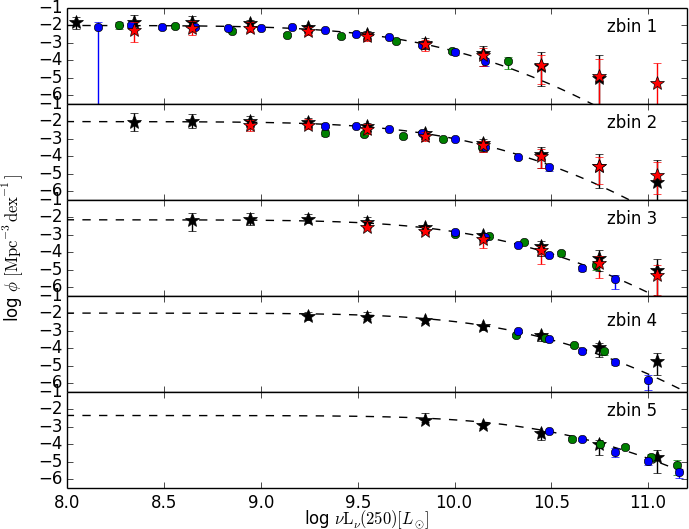}
\caption{ 250 $\mu$m LF in five redshift bins. Our results are plotted as filled stars (black: Stripe 82; red: GAMA fields), which agree well with previous measurements (green circles: Dye et al.  2010; blue circles: Marchetti et al.  2015). The dashed line is the best fit to our measurements (GAMA fields and Stripe 82) and Marchetti et al. (2015).}
\label{LF_zbins}
\end{figure}

Fig. ~\ref{mean_property} shows the mean rest-frame 250 $\mu$m luminosity $L_{\rm 250}$ as a function of $M_r$ for the DR 12 and Stripe 82 galaxies. There is good agreement in the overlapping $M_r$ range; this agreement is generally below 0.1 dex difference. At the faint end, galaxies exhibit a steep correlation between $L_{\rm 250}$ and $M_r$ without significant evolution with redshift.  At the bright end, the mean $L_{\rm 250}$ as a function of $M_r$ begins to flatten with significant redshift evolution. As optically red galaxies dominate at the bright end, the redshift evolution can be explained by the evolution in the red galaxy population, which was first observed in Bourne et al. (2012). In the two highest redshift bins, $z4$ and $z5$, the depth of DR 12 means that we are only able to probe the bright galaxies with a flattened relation between the mean $L_{\rm 250}$ and $M_r$. As explained in Appendix A, our method only works if there is a roughly monotonic relation between the mean $L_{\rm 250}$ and $M_r$. Therefore, we do not use DR 12 at $z>0.3$. Fig.~\ref{LF_zbins} shows our reconstructed rest-frame 250 $\mu$m LF, using DR 12 in the GAMA fields and the deeper Coadd data in Stripe 82. The luminosity limit reached by our method corresponds to the mean $L_{250}$ of the galaxies in the faintest $M_r$ bin in each redshift slice. Good agreement can be found between our results and previous determinations in the overlapping luminosity range. The dashed line in each panel is a modified Schechter function (Saunders et al. 1990) fit to our results (in the GAMA fields and Stripe 82) and measurements from Marchetti et al. (2015), 
\begin{equation}
\phi(L) = \frac{dn}{dL}=\phi^* \left(\frac{L}{L^*}\right)^{1-\alpha} \exp{\left[-\frac{1}{2\sigma^2} \log^2_{10}\left(1+\frac{L}{L^*}\right)\right]},
\end{equation}
where $\phi^*$ is the characteristic density, $L^*$ is the characteristic luminosity, $\alpha$ describes the faint-end slope, and $\sigma$ controls the shape of the cut-off at the bright end. We assume $\sigma$ and $\alpha$ do not change with redshift. Table 1 lists the best-fit and marginalised error for the parameters in the modified Schechter function. We find strong positive luminosity evolution $L^*_{250}(z)\propto(1+z)^{4.89\pm1.07}$ and moderate negative density evolution $\Phi^*_{250}(z)\propto(1+z)^{-1.02\pm0.54}$ over $0.02 < z< 0.5$.

\begin{table}
\caption{Best-fit values and marginalised errors of the parameters in the modified Schechter functions.}\label{table:selection}
\begin{tabular}{lll}
Parameter & Best  value &  error  \\
\hline
$\log L_1^*$ ($z1=[0.02, 0.1]$) &       9.17   &  0.11   \\
\hline
$\log L_2^*$ ($z2=[0.1, 0.2]$) &      9.37   &  0.11 \\
\hline
$\log L_3^*$ ($z3=[0.2, 0.3]$) &   9.50  &   0.12  \\
\hline
$\log L_4^*$ ($z4=[0.3, 0.4]$) &   9.66    & 0.12  \\
\hline
$\log L_5^*$ ($z5=[0.4, 0.5]$) &   9.87    & 0.13  \\
\hline
$\log \phi_1^*$ ($z1=[0.02, 0.1]$) &  -1.60   & 0.02  \\
\hline
$\log \phi_2^*$ ($z2=[0.1, 0.2]$) &  -1.60    & 0.03 \\
\hline
$\log \phi_3^*$ ($z3=[0.2, 0.3]$) &  -1.70  &  0.06  \\
\hline
$\log \phi_4^*$ ($z4=[0.3, 0.4]$) &  -1.59   &  0.10  \\
\hline
$\log \phi_5^*$ ($z5=[0.4, 0.5]$) &  -1.92   &  0.13  \\
\hline
$\sigma$&  0.35  &  0.01  \\
\hline
$\alpha$&  1.03 & 0.02  \\
\hline
\end{tabular}
\end{table}

\section{Conclusion}
 
 We study the low-redshift, rest-frame 250 $\mu$m LF using stacking of deep optically selected galaxies from the SDSS survey on the {\it Herschel}-SPIRE maps of the GAMA fields and the Stripe 82 area. Our method not only recovers the mean 250 $\mu$m luminosities $L_{250}$ of galaxies that are too faint to be individually detected, but also their underlying distribution functions.We find very good agreement with previous measurements. More importantly, our stacking method probes the LF down to much fainter luminosities ($\sim25$ times fainter) than achieved by previous efforts. We find strong positive luminosity evolution $L^*_{250}(z)\propto(1+z)^{4.89\pm1.07}$ and moderate negative density evolution $\Phi^*_{250}(z)\propto(1+z)^{-1.02\pm0.54}$ at $z< 0.5$. Our method bypasses some major difficulties in previous studies, however, it critically relies on the input photometric redshift catalogue. Therefore, issues such as photometric redshift bias and accuracy would have an impact. Over the coming years, our stacking method of reconstructing the LF will deliver even more accurate results and also extend to even fainter luminosities and higher redshifts. This is because, although we are probably not going to have any FIR/sub-mm imaging facility that will surpass {\it Herschel} in terms of areal coverage, sensitivity, and resolution in the near future, our knowledge of the optical and near-IR Universe will increase dramatically with ongoing and planned surveys such as DES and LSST. In addition, large and deep spectroscopic surveys such as EUCLID and DESI will further improve the quality of photometric redshift.

\begin{appendix}

\section{The modified stacking method}

Below we summarise the main steps of recovering the 250 $\mu$m LF in a given redshift bin using our modified stacking method:

1. Stack the 250 $\mu$m images centred on the galaxies in a given $M_r$ bin, weighted by luminosity distance squared ($D_L^2$) and $k-$correction. Measure the mean and standard deviation of the rest-frame 250 $\mu$m luminosity $L_{\rm 250}$, i.e. $\bar{m}$ and $\bar{s}$. Note that the estimates $\bar{m}$ and $\bar{s}$ are affected by instrument noise in the 250 $\mu$m images.

2. Generate $n$ bootstrap realisations for each sample (i.e. the set of galaxies in a given $M_r$ bin) and repeat Step 1 for all realisations. Form an estimate of the error on $\bar{m}$ and $\bar{s}$, using the $n$ bootstrap realisations.

3. Generate synthetic galaxies\footnote{The number of synthetic galaxies is equal to the number of real galaxies in a given $z$ and $M_r$ bin.} with random $L_{\rm 250}$ values drawn from log-normal distributions set by known $\mu$ and $\sigma$ values and add them to random locations in the map.  The $\sigma$ values (i.e. the standard deviation of $\log L_{\rm IR}$) are chosen to sample linearly between 0.027 and 2.17 with a width of 0.027. The $\mu$ values  (i.e. the mean of $\log L_{\rm IR}$) are sampled linearly between 6.478 and 12.088 with a width of 0.035. Measure the mean and standard deviation of $L_{\rm 250}$ of the synthetic galaxies, taking into account the effect of instrument noise in the 250 $\mu$m images.

4. Repeat Step 3 $n$ times. Each time sampling different random locations in the maps.

5. By comparing the measured mean and standard deviation of $L_{\rm 250}$ of the real galaxies with the mean and standard deviation estimates of the synthetic galaxies (for all $n$ repetitions), select all sets of $\mu$ and $\sigma$ values that give reasonably close mean and standard deviation to the real values using $\chi^2$ statistics.

6. For each set from the accepted $\mu$ and $\sigma$ values, generate log-normally distributed $L_{\rm 250}$ and assign them randomly to galaxies in a given $M_r$ bin. Calculate the resulting distribution function of $L_{\rm 250}$.

7. Repeat Step 6 for all accepted values of $\mu$ and $\sigma$, so we have multiple realisations of the distribution function of $L_{\rm 250}$ for galaxies in a single $M_r$ bin.

8. Repeat Step 1 to 7 for all $M_r$ bins. The 250 $\mu$m LF is derived by adding up contributions to a given $L_{\rm 250}$ bin from galaxies in all $M_r$ bins. Using the multiple realisations, form a median estimate of the final 250 $\mu$m LF and its confidence range.

For this method to work properly, it is important that the mean r-band luminosity $M_r$ and the mean 250 $\mu$m luminosity has a more or less monotonic relation. Otherwise, one could have situations where some sources in a given bin in $L_{250}$ have fainter $M_r$ values than are included in the optical prior list. Our method is similar to the stacking approach in Bethermin et al. (2012) which was used to derive the SPIRE number counts. The main difference is that, in Bethermin et al. (2012), the aim was to recover the mean and standard deviation of the logarithm of the 250 $\mu$m flux rather than luminosity. 

\end{appendix}

\begin{acknowledgements}

PN acknowledges the support of the Royal Society through the award of a University Research Fellowship, the European Research Council, through receipt of a Starting Grant (DEGAS-259586) and the support of the Science and Technology Facilities Council (ST/L00075X/1). NB acknowledges funding from the European Union Seventh Framework Programme (FP7/2007-2013) under grant agreement no. 312725. LD and SJM acknowledge support from the European Research Council Advanced Investigator grant, COSMICISM and Consolidator grant, cosmic dust.

The H-ATLAS is a project with Herschel, which is an ESA space observatory with science instruments provided by European-led Principal Investigator consortia and with important participation from NASA. The H-ATLAS web site is \url{http://www.h-atlas.org/}.

Funding for SDSS-III has been provided by the Alfred P. Sloan Foundation, the Participating Institutions, the National Science Foundation, and the U.S. Department of Energy Office of Science. The SDSS-III web site is http://www.sdss3.org/.

SDSS-III is managed by the Astrophysical Research Consortium for the Participating Institutions of the SDSS-III Collaboration including the University of Arizona, the Brazilian Participation Group, Brookhaven National Laboratory, Carnegie Mellon University, University of Florida, the French Participation Group, the German Participation Group, Harvard University, the Instituto de Astrofisica de Canarias, the Michigan State/Notre Dame/JINA Participation Group, Johns Hopkins University, Lawrence Berkeley National Laboratory, Max Planck Institute for Astrophysics, Max Planck Institute for Extraterrestrial Physics, New Mexico State University, New York University, Ohio State University, Pennsylvania State University, University of Portsmouth, Princeton University, the Spanish Participation Group, University of Tokyo, University of Utah, Vanderbilt University, University of Virginia, University of Washington, and Yale University.
\end{acknowledgements}

\end{document}